\documentclass[11pt]{article}
\usepackage{times}
\usepackage{mathfont} 
\usepackage{cite}
\usepackage{url}

\def\log{\mathop{\rm log}}

\DeclareSymbolFont{AMSb}{U}{msb}{m}{n}
\DeclareSymbolFontAlphabet{\Bbb}{AMSb}

\DeclareSymbolFont{lasy}{U}{lasy}{m}{n}
\let\Box\undefined
\DeclareMathSymbol\Box{0}{lasy}{"32}

\newtheorem{lemma}{Lemma}

\newtheorem{theorem}{Theorem}

\newcommand{\qed}{~$\Box$\medbreak}
\newenvironment{proof}{\noindent{\bf Proof: }}{\qed}

\input epsf
\def\efig#1#2{\hbox{\epsfxsize=#1\epsfbox{#2}}}

\def\dg{^\circ}

\newcounter{bibnos}
\def\thebibliography#1{\list
  {[\arabic{enumi}]}{\settowidth\labelwidth{[100]}\leftmargin\labelwidth
  \advance\leftmargin\labelsep
  \usecounter{enumi}}%
  \setcounter{enumi}{\thebibnos}%
  \def\newblock{\hskip .11em plus .33em minus .07em}%
  \sloppy\clubpenalty4000\widowpenalty4000%
  \itemsep-2pt%
  \sfcode`\.=1000\relax}

\begin{document}
\bibliographystyle{abuser}

\title{Quadrilateral Meshing by Circle Packing}

\author{Marshall Bern~\thanks{Xerox Palo Alto Research Center, 3333 Coyote
Hill Road, Palo Alto, CA, 94304, bern@parc.xerox.com.}
\and
David Eppstein~\thanks{Dept.
Information and Computer Science, Univ. of California, Irvine, CA
92697-3425,
eppstein@ics.uci.edu.}}

\date{}
\maketitle
\thispagestyle{empty}

\begin{abstract}
We use circle-packing methods to generate quadrilateral meshes for
polygonal domains, with guaranteed bounds both on the quality and the
number of elements.  We show that these methods can generate meshes of
several types: (1) the elements form the cells of a Vorono{\"\i} diagram,
(2) all elements have two opposite $90\dg$ angles, (3) all elements
are kites, or (4) all angles are at most $120\dg$. In each case the
total number of elements is $O(n)$, where $n$ is the number of
input vertices.
\end{abstract} 

\section{Introduction}

We investigate here problems of unstructured quadrilateral mesh
generation for polygonal domains, with two conflicting requirements.
First, we require there to be few quadrilaterals, linear in the number of
input vertices; this is appropriate for methods in which high order
basis functions are used, or in multiblock grid generation in which each
quadrilateral is to be further subdivided into a structured mesh. Second,
we require some guarantees on the quality of the mesh: either the
elements themselves should have shapes restricted to certain classes of
quadrilaterals, or the mesh should satisfy some more global quality
requirements.

Computing a linear-size quadrilateralization, without regard for quality,
is quite easy. One can find quadrilateral meshes with few elements, for
instance, by triangulating the domain and subdividing each triangle into
three quadrilaterals\cite{RamRamTou-CCCG-95}. For convex domains, it is
possible to exactly minimize the number of elements\cite{MueWei-SCG-97}.
However these methods may produce very poor quality meshes. High-quality
quadrilateralization, without rigorous bounds on the number of elements,
is an area of active practical interest. Techniques such as
paving\cite{BlaSte-IJNME-91} can generate high-quality meshes for typical
inputs; however these meshes may have many more than $O(n)$ elements.
Indeed, if the requirements on element quality include a constant bound
on aspect ratio, then meshing a rectangle of aspect ratio $A$ will require
$\Omega(A)$ quadrilaterals, even though in this case $n=4$.

We provide a first investigation into the problem of finding
a suitable tradeoff between those two requirements: for which measures of
mesh quality is it possible to find guaranteed-quality meshes with
guaranteed linear complexity? The results---and indeed the
algorithms---of this paper are analogous to the problem of nonobtuse
triangulation\cite{BakGroRaf-DCG-88,BerEpp-IJCGA-92,BerMitRup-DCG-95}.
Interestingly, for quadrilaterals there seem to be several analogues of
nonobtuseness.

\looseness=-1
Our algorithms are based on {\em circle packing}, a powerful geometric
technique useful in a variety of contexts. 
Specifically, we build on a circle-packing method due to Bern,
Mitchell, and Ruppert\cite{BerMitRup-DCG-95}.  In this method, before
constructing a mesh, one fills the domain with circles, packed closely
together so that the gaps between them are surrounded by three or four
tangent circles. (Circle packings with only three-sided gaps form
a sort of discrete analogue to conformal mappings\cite{Ste-CMFT-97}.
However for most domains some four-sided gaps are necessary, and in some
of our algorithms four-sided gaps are actually helpful since they lead to
degree-four mesh vertices.) One then uses
these circles as a framework to construct the mesh, by placing mesh
vertices at circle centers, points of tangency, and within each gap.
Earlier work by Shimada and Gossard\cite{ShiGos-SMA-95} also uses
approximate circle packings and sphere packings to
construct triangular meshes of 2-d domains and 3-d surfaces. Other
authors have introduced related circle packing ideas into meshing via
conforming Delaunay triangulation\cite{NacSri-CCCG-91}, conformal
mapping\cite{DoyHeRod-DCG-94,RodSul-JDG-87}, and
decimation\cite{MilTalTen-IMR-96,MilTalTen-SODA-97}.
Circle packing ideas closely related to the algorithms in this paper
have also been applied in origami
design\cite{Lan-SCG-96,BerDemEpp-FUN-98}.

\begin{figure}[t]
$$\efig{5in}{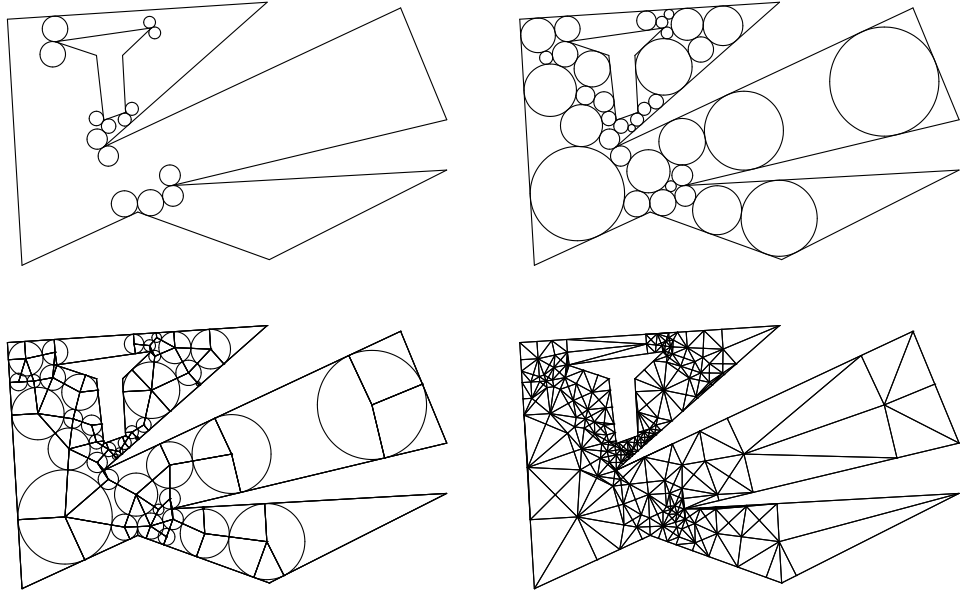}$$
\caption{Nonobtuse triangulation steps: (a) protect reflex vertices
and connect holes; (b) pack polygon with circles; (c) connect circle
centers; (d) triangulate remaining polygonal regions.}
\label{bmr}
\end{figure}

We use circle packing to develop four new quadrilateral meshing
methods.  First, in Section~\ref{sec:vor}, we show that the
Vorono{\"\i} diagram of the points of tangency of a suitable circle
packing forms a quadrilateral mesh. Although the individual elements in
this mesh may not have good quality, the Vorono{\"\i} structure of the
mesh may prove useful in some applications such as finite volume methods.
Second, in Section~\ref{sec:2rt}, we overlay this Vorono{\"\i} mesh
with its dual Delaunay triangulation; this overlay subdivides each
Vorono{\"\i} cell into quadrilaterals having two opposite right angles.
Note that any such quadrilateral must have all four of its vertices on a
common circle. Third, in Section~\ref{sec:kite}, we
show that a small change to the method of Bern et al.
(basically, omitting some edges), produces a mesh of {\em kites}
(quadrilaterals having two adjacent pairs of equal-length sides). The
resulting mesh optimizes the {\em cross ratio} of the elements (a measure
of the aspect ratio of the rectangles into which each element may be
conformally mapped): any kite can be conformally mapped onto a square.
Finally, in Section~\ref{sec:120}, we subdivide these kites into smaller
quadrilaterals, producing a mesh in which each quadrilateral has maximum
angle at most $120\dg$.  This is optimal: there exist domains for which
no mesh has angles better than
$120\dg$.

\begin{figure}[t]
$$\efig{3.5in}{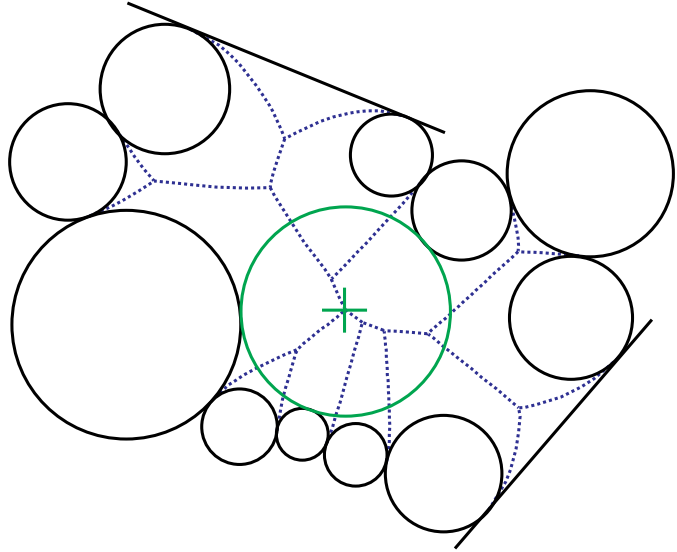}$$
\caption{Placement of a new circle centered on a Vorono{\"\i} vertex
partitions a region bounded by circular arcs into several simpler
regions.}
\label{arc-voronoi}
\end{figure}

\section{Circle Packing}

Let us first review the nonobtuse triangulation method of Bern et
al.\cite{BerMitRup-DCG-95}.  This algorithm is given an $n$-vertex
polygonal region (possibly with holes), and outputs a triangulation with
$O(n)$ new Steiner points in which no triangle has an obtuse angle.  In
outline, it performs the following steps:
\begin{enumerate}
\item {\em Protect reflex vertices} of the polygon by placing
circles tangent to the boundary of the polygon on
either side of them, small enough that they do not intersect each other or
other features of the polygon
(Figure~\ref{bmr}(a)).
\item {\em Connect holes} of the polygon by placing
nonoverlapping circles, tangent to edges of the polygon or to previously placed
circles, so that the domain outside the
circles forms one or more simply connected regions with circular-arc
sides.
\item {\em Simplify} each region by packing it with further circles until each
remaining region has three or four circular-arc or straight-line sides
(Figure~\ref{bmr}(b)).
\item {\em Partition} the polygon into 3- and 4-sided polygonal
regions by connecting the centers of tangent circles
(Figure~\ref{bmr}(c)).
\item {\em Triangulate} each region with nonobtuse triangles
(Figure~\ref{bmr}(d)).
\end{enumerate}

Our quadrilateralization algorithms will be based on the same general
outline, and in several cases the quadrilaterals we form can be viewed
as combinations of several of the triangles formed by this algorithm.

Steps 1, 4, and 5 are straightforward to implement.
Eppstein\cite{Epp-IJCGA-97} showed that step 2 could be implemented
efficiently, in time $O(n\log n)$, as independently did Mike Goodrich
and Roberto Tamassia, and Warren Smith (unpublished). We now describe in
some more detail step 3, simplification of regions, as we will need to
modify this step in some of our algorithms.

\begin{lemma}[Bern et al.\cite{BerMitRup-DCG-95}]
Any simply connected region of the plane bounded by $n$ circular arcs and
straight line segments, meeting at points of tangency, can be packed with
$O(n)$ additional circles in $O(n\log n)$ time, such that the remaining
regions between circles are bounded by at most four tangent circular arcs.
\end{lemma}

\begin{proof}
Compute the {\em Vorono{\"\i}
diagram} of the circles within this region; that is, a partition of the
region into cells, each of which contains points closer to one of the
circles than to any other circle (Figure~\ref{arc-voronoi}).  Because the
region is simply connected, the cell boundaries of this diagram form a
tree. We choose a vertex $v$ of this tree such that each of the subtrees
rooted at $v$ has at most half the leaves of the overall tree, and draw
a circle centered at $v$ and tangent to the circles having Vorono{\"\i}
cells incident at~$v$.  This splits the region into simpler regions.
We continue recursively within these regions, stopping when we reach
regions bounded by only four arcs (in which no further simplification is
possible).  Adding each new circle to the Vorono{\"\i} diagram
can be done in time linear in the number of arcs bounding the region, so
the total time to subdivide all regions is $O(n\log n)$.
\end{proof}

We call the region between circles of this packing a {\em gap}.
We now state without proof two technical results of Bern et al. about
these gaps.

\begin{lemma}[Bern et al.\cite{BerMitRup-DCG-95}]
The points of tangency on the boundary of a gap are cocircular.
\end{lemma}

\begin{figure}[t]
$$\efig{2.7in}{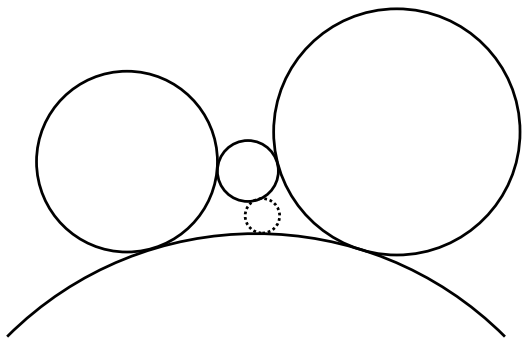}$$
\caption{Bad four-sided gap split into two good gaps.}
\label{bad-gap}
\end{figure}

A {\em three-sided gap} is one bounded by three circular arcs.
A {\em good four-sided gap} is a gap bounded by four arcs, such that the
circumcenter of its points of tangency is contained within the convex
hull of those points.  A {\em bad four-sided gap} is any other four-arc
gap.

\begin{lemma}[Bern et al.\cite{BerMitRup-DCG-95}]
\label{split-bad-gap}
Any bad four-sided gap can be split into two good four-sided gaps
by the
addition of a circle tangent to two of the bad four-sided gap's circles.
(Figure~\ref{bad-gap}.)
\end{lemma}

In some cases two opposite circles bounding one of the new gaps created by
Lemma~\ref{split-bad-gap} may overlap, but this poses no problem for the
rest of the algorithm.

\section{Vorono{\"\i} Quadrilateralization}
\label{sec:vor}

\begin{figure}[t]
$$\efig{4.5in}{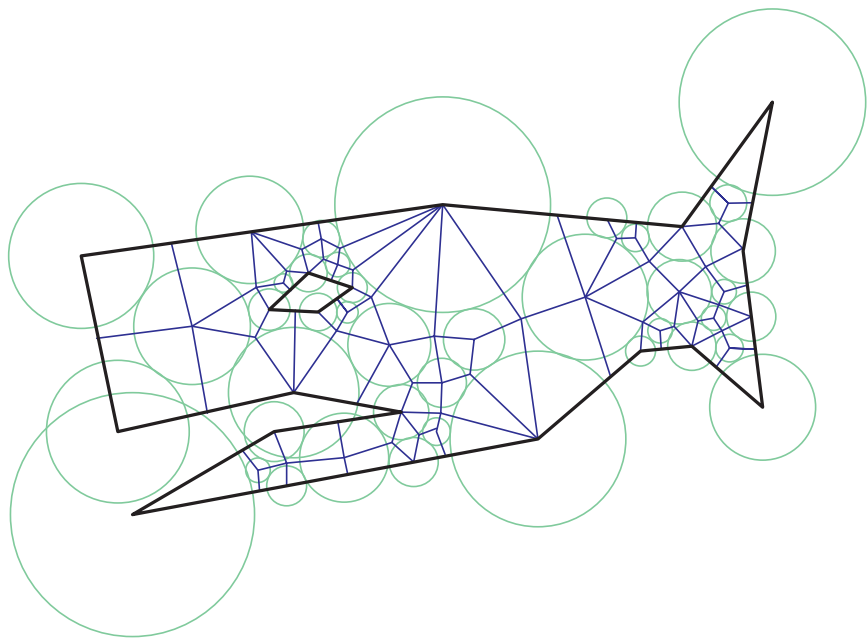}$$
\caption{Vorono{\"\i} quadrilateralization of a polygon.}
\label{whale}
\end{figure}

We begin with the quadrilateralization procedure most likely
to be useful in practice, due to its low output complexity and lack
of complicated special cases.

The {\em geodesic Vorono{\"\i} diagram} of a set of point {\em sites} in a
polygonal domain is a partition of the domain into cells, in each of
which the geodesic distance (distance along paths within the domain) is
closest to one of the given sites.  We now describe a method of finding
a point set for which the geodesic Vorono{\"\i} diagram forms a
quadrilateral mesh.  One potential application of this type of mesh
would be in the finite volume method, as the dual of this Vorono{\"\i}
mesh could be used to define control volumes for that method (see
e.g. Miller et al.\cite{MilTalTen-IMR-96}). The angle between each primal
and dual edge pair would be $90\dg$, causing some terms in the finite
volume method to cancel and therefore saving some
multiplications\cite{BakGroRaf-DCG-88,BerGil-IPL-92}.
Our mesh will also have the possibly useful property 
that all Vorono{\"\i} edges will cross their duals.

We modify the initial circle packing of Bern et
al.\cite{BerMitRup-DCG-95}, as follows.  We start by protecting vertices,
as before; but in this case that protection consists of a circle centered
at each domain vertex.  Then, as before we fill the remainder of the
domain by tangent circles; however we do not attempt to create tangencies
with the domain boundary; instead the circle packing should meet the
boundary at circles with their centers on the boundary. Further, no
tangent point between two circles should lie on the domain boundary,
although circles centered on the boundary may meet in the domain
interior.  (Some circles may cross
or be tangent to the boundary, however we ignore these incidences,
instead treating these circles as part of three-sided gaps.)
It is not hard to modify the previous circle packing
algorithms to meet these conditions.  The result will be a packing with,
again, three-sided and four-sided gaps.
However, the gaps involving boundary edges are all four-sided and
have right-angled corners rather than points of tangency on those edges.

\begin{figure}[t]
$$\efig{3.3in}{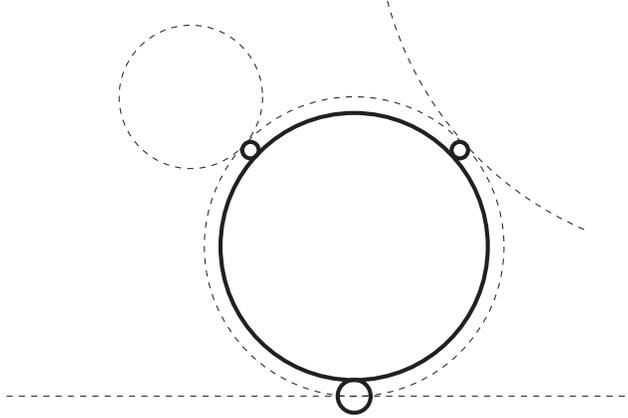}$$
\caption{Replacement of disk tangent to boundary by multiple disks
centered on boundary.}
\label{inset}
\end{figure}

\begin{theorem}\label{Voronoi}
In $O(n\log n)$ time we can find a circle packing as above,
such that the geodesic Vorono{\"\i} diagram of the points of tangencies of
the circles forms a quadrilateral mesh.
\end{theorem}

\begin{proof}
The vertex protection step can be performed in $O(n)$ time using
circles with radius half the minimum distance between vertices.
(This minimum distance is an edge of the Delaunay triangulation and can be
found in $O(n\log n)$ time.)
Next we find a set of $O(n)$
circles to add to our packing, so that any remaining gaps are simply
connected\cite{Epp-IJCGA-97}. If in this step we
ever add a circle $c$ tangent to the boundary, we replace it by a set of
circles: a circle with radius $\epsilon$ centered on the boundary at the
point of tangency, circles with radius $\epsilon/2$ tangent to $c$ at
each of its other points of tangency, and one circle concentric to $c$
with radius reduced by $\epsilon$; this replacement is depicted in
Figure~\ref{inset}.

Finally, as in the algorithm of Bern et al.\cite{BerMitRup-DCG-95} we
repeatedly find the Vorono{\"\i} diagram of the circles bounding any
remaining gap and place a circle on a Vorono{\"\i} vertex so as to divide
the gap into two smaller parts of roughly equal complexity.  However, in
order to avoid placing circles tangent to the boundary in this step, we
change their method by using only the circles around a gap as
Vorono{\"\i} sites, omitting any diagram edges that may bound the gap.
Our Voronoi diagram's edges (together with any boundary edges of the
gap) form a tree, so we can find a vertex which splits the tree's leaves
roughly evenly.  Placing a circle at that vertex produces two simpler
gaps and does not cause essential tangencies with the domain boundary.
Unlike Bern et al.\cite{BerMitRup-DCG-95}, we do not bother eliminating
bad four-sided gaps.

We form a mesh by connecting each center of a
circle in the packing to the circumcenters of adjacent gaps
(Figure~\ref{whale}).
In the four-sided gaps along the domain boundary,
we place an additional edge from the boundary to the center of the
opposite circle, bisecting the chord between the tangencies with the two
other circles.
These edges form a quadrilateral mesh since each face
surrounds a point of tangency, and each point of tangency is surrounded
by the vertices from two circles and two gaps.  Each mesh element is the
Voronoi cell of the point of tangency it contains; its boundary is
composed of perpendicular bisectors of dual Delaunay edges.  Each dual
Delaunay edge has a circumscribing circle from the circle packing as
witness to the empty-circle property of Delaunay graphs.
\end{proof}

Curiously, this mesh is not only a certain type of
generalized Vorono{\"\i} diagram; it is also another type of generalized
Delaunay triangulation!  The {\em power} of a circle with
respect to a point in the plane is the squared radius of the
circle minus the squared distance of the point to the circle's center.
The {\em power diagram} of a set of (not necessarily disjoint) circles is
a partition of the plane into cells, each consisting of the points for
which the power of some particular circle is greatest.  Like the usual
kind of Vorono{\"\i} diagram, the power diagram has convex cells, since
the separator between any two circles' cells is a line (if the two circles
overlap, their separator is the line through their two
intersection points).  We can restrict the power diagram to a polygonal
domain by defining the power only for points visible to the center of the
given circle.

\begin{figure}[t]
$$\efig{4.2in}{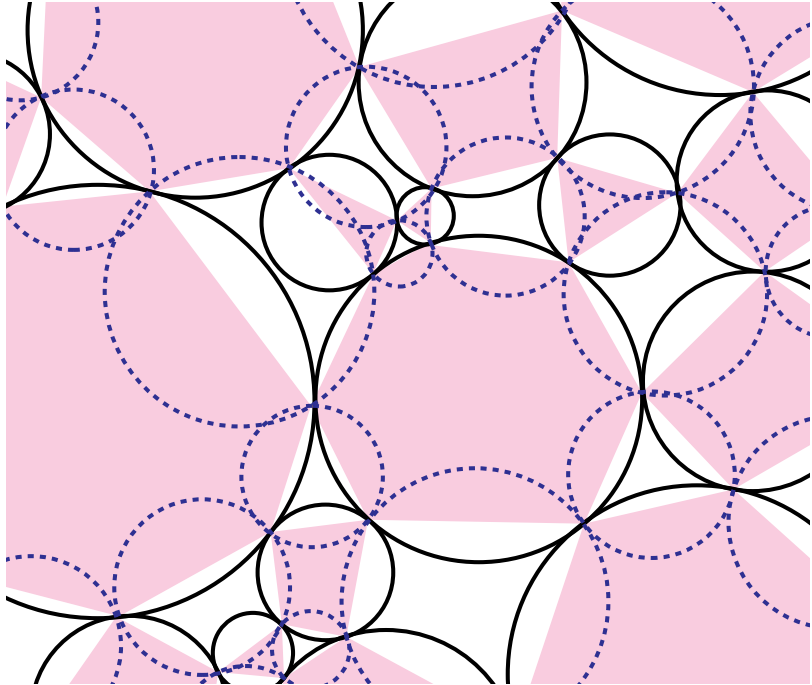}$$
\caption{Circle packing (solid circles), circumcircles of gaps (dashed
circles), and power diagram (shaded and unshaded polygons).}
\label{powerdiag}
\end{figure}

From the construction above, define a family $F$ of circles by
including the original packing and a ``dual'' collection of circles
through the tangencies surrounding each gap, centered at the gap's site.
As we now show, the power diagram of this family
(depicted in Figure~\ref{powerdiag}) is the planar dual to our mesh.

\begin{theorem}
The quadrilateral mesh defined above includes an edge between two points
if and only if the corresponding circles' cells share an edge in the power
diagram of $F$.
\end{theorem}

\begin{proof}
{}$F$ has one circle centered at each vertex; the two
circles corresponding to the endpoints of an edge overlap in a lune.
The lune's two corners are points of tangency in the original
circle packing (or, if the edge is on the
domain boundary, the corners are one such point of tangency and
its reflection) and are contained in the two quadrilaterals on either
side of the edge. These corners have power zero with respect to the two
circles, and are not interior to any other circles; therefore they have
those two circles (and possibly some others) as nearest power neighbors.
Since power diagram cells are convex, those two circles must continue to
be the nearest neighbors to each point along the center line of the lune;
in other words this center line lies along an edge in the power diagram
corresponding to the given mesh edge.

Conversely, we must show that every power diagram adjacency corresponds
to a mesh edge.  But the power diagram boundaries described above form a
convex polygon completely containing the center of the cell's circle;
therefore there can be no other adjacencies than the ones we have
already found, which correspond to mesh edges.
\end{proof}

Since quadrilaterals in this mesh typically correspond to eight
triangles in the nonobtuse triangulation algorithm of Bern et al.,
the constant factors in the $O(n)$ bound above should be quite small in
practice.  Bern et al.\cite{BerMitRup-DCG-95} observed that their method
typically generated between $20n$ and $30n$ triangles, so we should
expect between
$3n$ and
$4n$ quadrilaterals in our mesh.

\section{Opposite Right Angles}
\label{sec:2rt}

As we now show, the Vorono{\"\i} triangulation above can be used to find
another quadrilateral mesh, in which each quadrilateral has two opposite
right angles.  Such a quadrilateral must be {\em cyclic} (having all
four vertices on a common circle); further, the circumcenter bisects the
diagonal connecting the two remaining vertices.

Our algorithm works by overlaying the power diagram defined above onto
the quadrilaterals of
Theorem~\ref{Voronoi}, resulting in their subdivision into smaller
quadrilaterals. In order to perform this subdivision, we may need to place
a few additional circles into our packing. On the boundary of the domain,
the gaps between circles will be formed by chains of three tangent
circles, the two ends of which are circles centered on the domain
boundary.  The center circle in this chain is allowed to cross the
boundary; we ignore this crossing.  Reflecting such a chain across the
domain boundary edge produces a four-sided gap
partially outside the domain; like Bern et al.\cite{BerMitRup-DCG-95} we
say that this gap is good or bad if the convex hull of its points of
tangency contains or doesn't contain their circumcenter respectively.  The
algorithm of this section requires these gaps to be good. As in the
method of Bern et al.\cite{BerMitRup-DCG-95}, any bad four-sided gap can
be subdivided into two good four-sided gaps by the addition of another
circle which by symmetry can be placed with its center on the domain
boundary.

\begin{theorem}
In $O(n\log n)$ time we can partition any polygon into a mesh
of $O(n)$ quadrilaterals, each having two opposite right angles.
\end{theorem}

\begin{proof}
We form the Vorono{\"\i} quadrilateralization of Theorem~\ref{Voronoi},
and subdivide each quadrilateral $Q$ into four smaller quadrilaterals by
dropping perpendiculars from the Vorono{\"\i} site contained in $Q$
to each of $Q$'s four sides.  On edges where two cells of the
Vorono{\"\i} quadrilateralization meet, the two perpendiculars end at a
common vertex because they are the two halves of a chord connecting two
tangent points on the same circle. For the same reason, each
perpendicular meets the edge to which it is perpendicular without
crossing any other cell boundaries first.
\end{proof}

The same procedure of dropping perpendiculars will work whenever we have
a Vorono{\"\i} diagram in which the site generating each cell can be
connected by a perpendicular to each cell edge.  Therefore, some
heuristic simplification can be applied to the mesh above, reducing its
complexity further: after forming the Vorono{\"\i} quadrilateralization of
Theorem~\ref{Voronoi}, remove sites one by one from the set of
generators as long as this condition is met.

\section{Kites}
\label{sec:kite}

\begin{figure}[t]
$$\efig{4.5in}{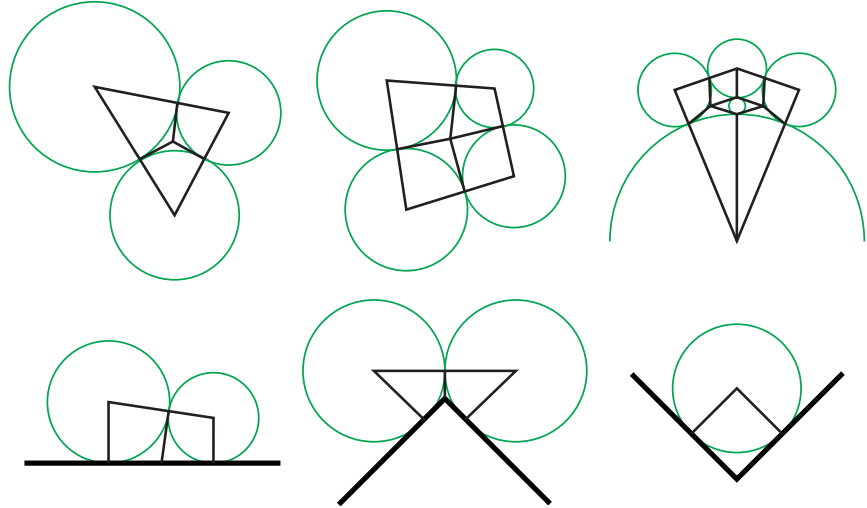}$$
\caption{Cases for decomposition into kites:
(a) three tangent circles; (b) four tangent circles forming good
four-sided gap; (c) bad four-sided gap subdivided into two good
four-sided gaps; (d) two tangent circles on boundary; (e) reflex vertex;
(f) convex vertex.}
\label{kite-cases}
\end{figure}

\begin{figure}[t]
$$\efig{4.7in}{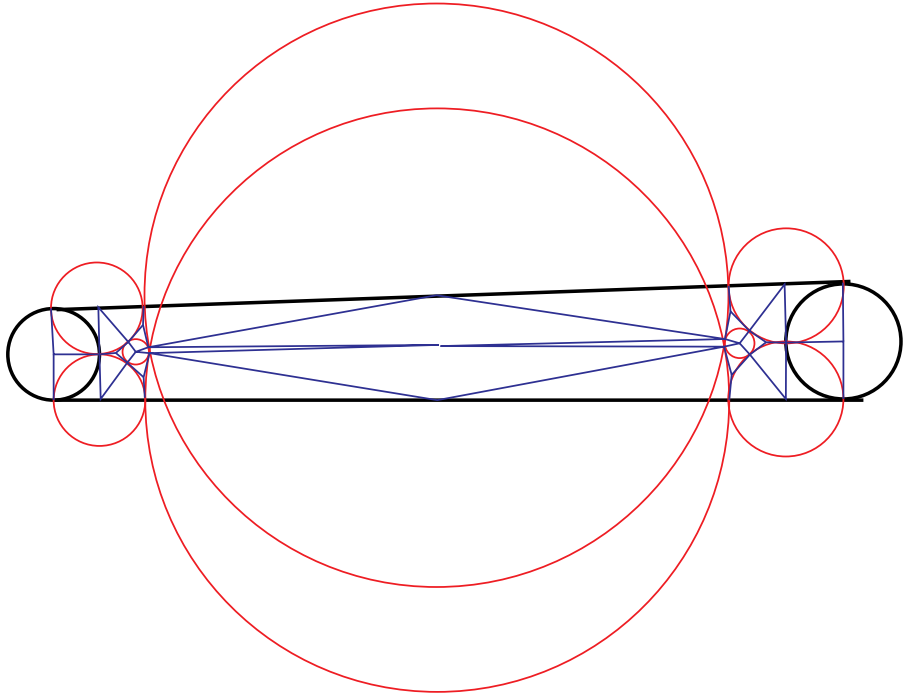}$$
\caption{Kite decomposition of four-sided gap with two sides on domain
boundary.}
\label{fixed-quad}
\end{figure}

\begin{figure}[t]
$$\efig{3in}{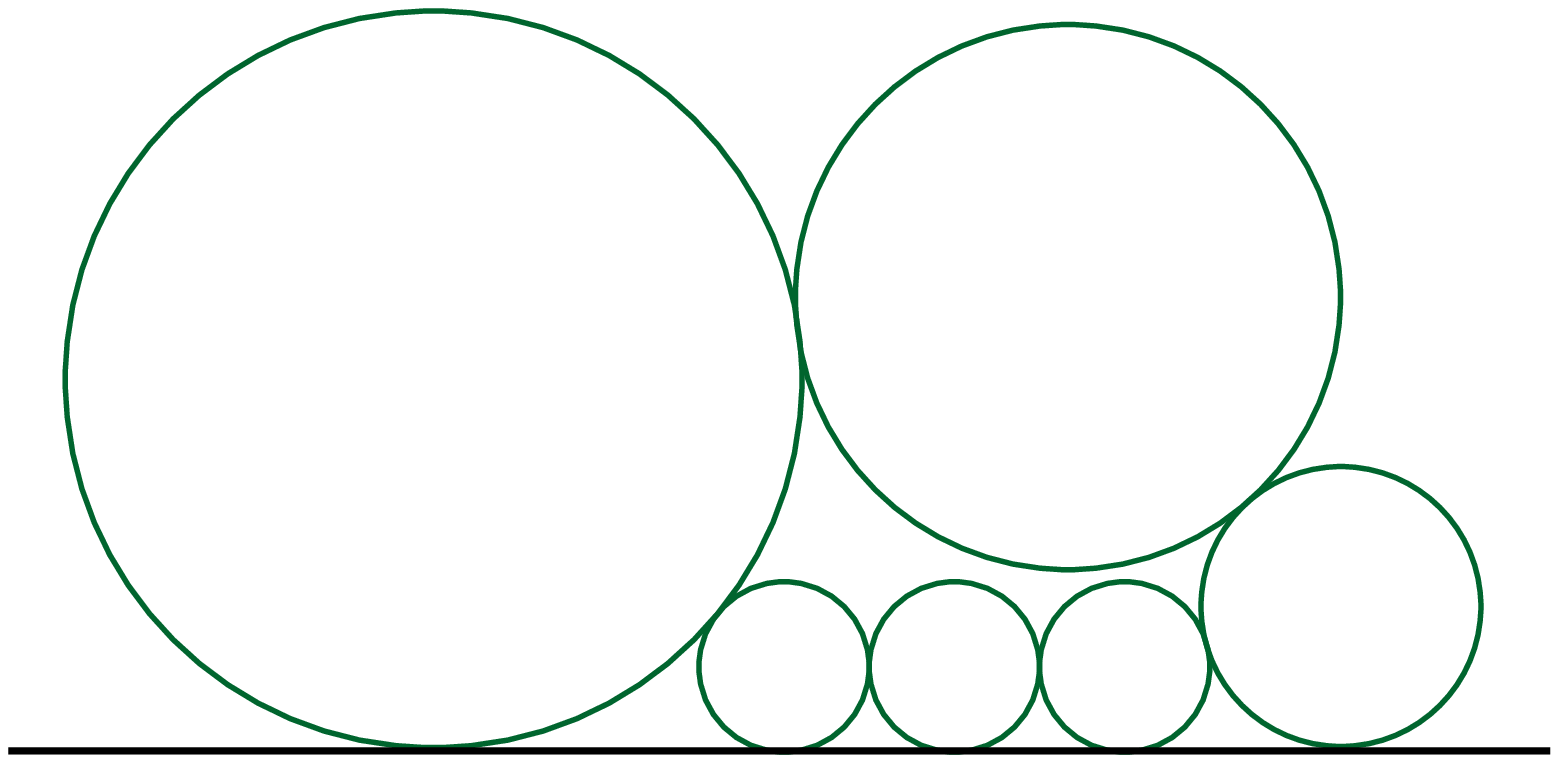}$$
\caption{Kite decomposition of four-sided gap with one side on domain
boundary: add small circles along boundary edge making three-sided gaps.}
\label{four-bdy}
\end{figure}

The next type of quadrilateralization we describe is one in which all
quadrilaterals are {\em kites} (convex quadrilaterals with an axis of
symmetry along one diagonal).  Although kites may have bad angles
(very close to $0\dg$ or $180\dg$), they have some other nice theoretical
properties. In particular, the {\em cross ratio} of a kite is always one.

The cross ratio of a quadrilateral with consecutive side lengths $a$, $b$,
$c$, and $d$ is the ratio $ac:bd$. Since this ratio is invariant under
conformal mappings, a conformal mapping from the quadrilateral to a
rectangle (taking vertices to vertices) can only exist if the
rectangle has the same cross ratio; but the cross ratio of a rectangle
is just the square of its aspect ratio.  Therefore, kites are among the
few quadrilaterals that can be conformally mapped onto squares.

\begin{theorem}\label{kites}
In $O(n\log n)$ time we can partition any polygon into a mesh
of $O(n)$ kites.
\end{theorem}

\begin{proof}
As in the algorithm of Bern et al., we find a circle packing; however as
discussed below we place some further constraints on the placement of
circles.  We then connect
pairs of tangent circles by radial line segments through their points of
tangency, and apply a case analysis to the resulting set of polygons.
As shown in Figure~\ref{kite-cases}, all interior gaps can be subdivided
into kites: three-sided gaps result in three kites, good four-sided gaps
result in four, and bad four-sided gaps result in seven.
Also shown in the figure are three types of gaps on the boundary of the
polygon: three-sided gaps along the edge, reflex vertices protected by
two equal tangent circles, and convex vertices packed by a single
circle.

There are two remaining cases, in which one or two of the sides of a
four-sided gap are portions of the domain boundary, and the
four-sided gap has a high aspect ratio preventing these boundary edges
from being covered by a small number of three-sided gaps.
In the simpler of these cases,
two opposite sides of the four-sided gap are both boundary edges.
Such a gap is necessarily good.  If it has aspect ratio $O(1)$, we can
line the domain edges by $O(1)$ additional circles, as in the next case.
Otherwise, our construction is illustrated in Figure~\ref{fixed-quad}.
We find a mesh using an auxiliary set of circles, perpendicular to the
original packing.  We first place at each end of the four-sided gap a
pair of identical circles, tangent to each other and crossing the
boundary edges perpendicularly at their points of tangency. These are the
medium-sized circles in the figure. We next place two more circles, each
perpendicular to one of the boundary edges and crossing it at the same
points already crossed by the previously added circles; these are the
large overlapping circles in the figure.  Finally, each end of the
original four-sided gap now contains a gap formed by four circles, but
two of these circles cross rather than sharing a tangency. We fill each
gap with an additional circle; these are the small circles in the
figure.  The resulting set of eight circles forms six three-sided gaps
and one good four-sided gap, and can be meshed as shown in the figure.

The final case consists of four-sided gaps (not necessarily
good) involving one boundary edge. To make this case tractable, we
restrict our initial placement of circles so that, if we place a circle
$C$ within a gap involving boundary edges, then $C$ is either tangent to
those edges or separated from them by a distance of at least $\epsilon$
times its radius, for some sufficiently small value~$\epsilon$.  Then,
any remaining four-sided boundary gap must have bounded aspect ratio, and
we can place
$O(1)$ small circles along the boundary edge leaving only three-sided
gaps on that edge (Figure~\ref{four-bdy}).  The interior of the gap can
then be packed with
$O(1)$ additional circles leaving only the previously solved three- and
four-sided internal gap cases.
\end{proof}

\section{No Large Angles}
\label{sec:120}

The maximum angle of any triangle has been shown to be one of the more
important indicators of triangular mesh quality\cite{BabAzi-SJNA-76},
and it is believed that the maximum angle is similarly important in
quadrilateral meshes.  For triangular meshes, a maximum angle of $90\dg$
can be achieved\cite{BerMitRup-DCG-95}, but for quadrilaterals this
would imply that all elements are rectangles, which can only be achieved
when the domain has axis-parallel sides.  Indeed, as we now show,
some domains require $120\dg$ angles.

\begin{theorem}
Any simple polygon with all angles at least $120\dg$
cannot be meshed by quadrilaterals having all angles less than $120\dg$.
\end{theorem}

\begin{proof}
Suppose we have such a simple polygon, and a quadrilateral mesh on it.
Let $x$ denote the number of mesh vertices on the boundary of the
polygon, $i$ denote the number of interior vertices, $e$ denote the
number of mesh edges, and $q$ denote the number of mesh quadrilaterals.
Then, since each quadrilateral has four edges, each interior edge
appears twice, and there are $x$ boundary edges, we have the relation
$4q=2e-x$.  Combining this with Euler's formula $x+i+q-e=1$ and
cancelling $q$ leaves $e=2i+(3/2)x-2$.
However, if all interior vertices of the mesh were incident to
four or more edges, and all exterior vertices were incident to
three or more edges, we would have $e\ge 2i + (3/2)x$ (since each edge
contributes two to the sum of vertex degrees), a contradiction.
So, the mesh has either an interior vertex with degree three, or an
exterior vertex with degree two, and in either case at least one of the
angles at that vertex must be at least $120\dg$.
\end{proof}

As we now show, this lower bound can be matched by our circle packing
methods.

\begin{figure}[t]
$$\efig{4in}{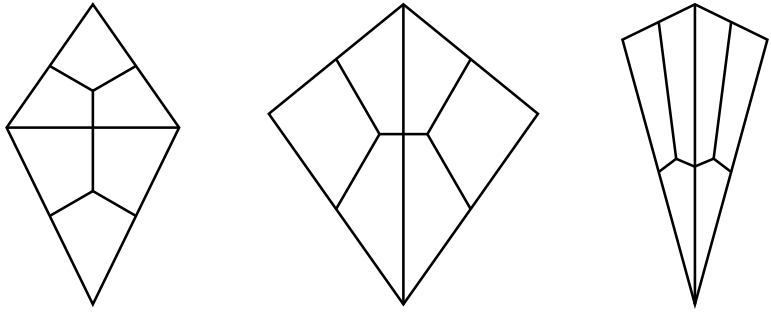}$$
\caption{Kites divided into six quadrilaterals with no angle
larger than $120\dg$: (a) top and bottom angles both less than $120\dg$;
(b) top and bottom angles both greater than $60\dg$; (c) top at least
$120\dg$ and bottom less than $120\dg$.}
\label{three-kites}
\end{figure}

\begin{theorem}
In $O(n\log n)$ time we can partition any polygon into a mesh
of $O(n)$ quadrilaterals with maximum angle~$120\dg$.
\end{theorem}

\begin{proof}
The result follows from Theorem~\ref{kites}, since any kite
(which we can assume without loss of generality to have a vertical axis
of symmetry) can be divided into six $120\dg$ quadrilaterals in one of
three ways depending on how the top and bottom angles of the kite compare
to $120\dg$.

Specifically, we add new subdivision points on the midpoints of each
kite edge.  Then, if both the top and bottom angle of the kite are sharp
(less than $120\dg$), we can split the kite along a line between the
left and right vertices, and subdivide both of the resulting triangles
into three $120\dg$ quadrilaterals (Figure~\ref{three-kites}(a)).
If both angles are large (greater than $60\dg$), we can similarly split
the kite vertically along a line from top to bottom and again subdivide
both of the resulting triangles (Figure~\ref{three-kites}(b)).
In both of these two cases the subdivisions are axis-aligned or at
$60\dg$ angles to the axes.  In the final case, the top angle is large
(at least $120\dg$) and the bottom is sharp (less than $120\dg$).
In this case, like the second, we partition the kite vertically into two
triangles, and again partition each triangle into three; however in this
final case the subdivisions are along lines between the bottom of the
triangle and the two opposite edge midpoints, and at $60\dg$ angles to
those lines.  It is easily verified that with the given assumptions on
the angles of the original kite, all vertices of the subdivision lie as
depicted in the figures and all angles are at most $120\dg$.
\end{proof}

\section{Conclusions}

We have shown that circle packing may be used in a variety of ways for
quadrilateral mesh generation with simultaneous guaranteed bounds on
complexity and quality.

Many questions remain open:
How small can we make the constant factors in our complexity bounds,
both in the worst case and in practice?
Can we generate
linear-complexity quadrilateral meshes with no small angles?
Can we combine guarantees on several quality measures at once?
Extensions of the circle packing method to three dimensional tetrahedral
or hexahedral meshing would be of interest, but seem difficult due to
the inability of three dimensional spheres to partition the domain into
bounded-complexity regions.  However perhaps our methods can be
generalized to guaranteed-quality quadrilateral surface meshes.

Some of the methods we describe are purely of theoretical interest, due
to high constant factors or distorted quadrilateral shapes, but we
believe circle packing should be useful in practice as well.
Among our methods, perhaps the low constant factors and lack
of complicated cases in the Vorono{\"\i}
quadrilateralization make it the most practical choice.

\section*{Acknowledgements}
Eppstein's work was supported in part by NSF grant CCR-9258355 and
by matching funds from Xerox Corp.

\vfill\eject
\section*{References}
\bibliography{cpack}

\begin{thebibliography}{10}

\bibitem{BabAzi-SJNA-76}
I.~Babu{\v{s}}ka and A.~Aziz.
\newblock {On the angle condition in the finite element method}.
\newblock {\em SIAM J. Numerical Analysis} 13:214--227, 1976.

\bibitem{BakGroRaf-DCG-88}
B.~S. Baker, E.~Grosse, and C.~S. Rafferty.
\newblock {Nonobtuse triangulation of polygons}.
\newblock {\em Discrete {\&} Computational Geometry} 3:147--168, 1988.

\bibitem{BerDemEpp-FUN-98}
M.~W. Bern, E.~Demaine, D.~Eppstein, and B.~Hayes.
\newblock {A disk-packing algorithm for an origami magic trick}.
\newblock {\em Proc. Int. Conf. Fun with Algorithms}, 1998.

\bibitem{BerEpp-IJCGA-92}
M.~W. Bern and D.~Eppstein.
\newblock {Polynomial-size nonobtuse triangulation of polygons}.
\newblock {\em Int. J. Computational Geometry {\&} Applications} 2:241--255,
  1992.

\bibitem{BerGil-IPL-92}
M.~W. Bern and J.~R. Gilbert.
\newblock {Drawing the planar dual}.
\newblock {\em Inf. Proc. Lett.} 43:7--13, 1992.

\bibitem{BerMitRup-DCG-95}
M.~W. Bern, S.~A. Mitchell, and J.~Ruppert.
\newblock {Linear-size nonobtuse triangulation of polygons}.
\newblock {\em Discrete {\&} Computational Geometry} 14:411--428, 1995.

\bibitem{BlaSte-IJNME-91}
T.~D. Blacker and M.~B. Stephenson.
\newblock {Paving: a new approach to automated quadrilateral mesh generation}.
\newblock {\em Int. J. Numer. Meth. Engr.} 32:811--847, 1991.

\bibitem{DoyHeRod-DCG-94}
P.~Doyle, Z.-X. He, and B.~Rodin.
\newblock {Second derivatives of circle packings and conformal mappings}.
\newblock {\em Discrete {\&} Computational Geometry} 11:35--49, 1994.

\bibitem{Epp-IJCGA-97}
D.~Eppstein.
\newblock {Faster circle packing with application to nonobtuse triangulation}.
\newblock {\em Int. J. Computational Geometry {\&} Applications} 7(5):485--491,
  1997.

\bibitem{Lan-SCG-96}
R.~Lang.
\newblock {A computational algorithm for origami design}.
\newblock {\em Proc. 12th Symp. Computational Geometry}, pp. 98--105. ACM,
  1996.

\bibitem{MilTalTen-SODA-97}
G.~L. Miller, D.~Talmor, and S.-H. Teng.
\newblock {Optimal good-aspect-ratio coarsening for unstructured meshes}.
\newblock {\em Proc. 8th Symp. Discrete Algorithms}, pp. 538--547. ACM and
  SIAM, January 1997.

\bibitem{MilTalTen-IMR-96}
G.~L. Miller, D.~Talmor, S.-H. Teng, N.~Walkington, and H.~Wang.
\newblock {Control volume meshes using sphere packing: generation, refinement
  and coarsening}.
\newblock {\em Proc. 5th Int. Meshing Roundtable}, pp. 47--61. Sandia National
  Laboratories, October 1996,
\url{http://sass577.endo.sandia.gov/9225/Personnel/samitch/roundtable96/papers/miller-final2631.ps.gz}.

\bibitem{MueWei-SCG-97}
M.~M{\"u}ller-Hannemann and K.~Weihe.
\newblock {Minimum strictly convex quadrangulations of convex polygons}.
\newblock {\em Proc. 13th Symp. Computational Geometry}, pp. 183--202. ACM,
  June 1997.

\bibitem{NacSri-CCCG-91}
L.~R. Nackman and V.~Srinivasan.
\newblock {Point placement for Delaunay triangulation of polygonal domains}.
\newblock {\em Proc. 3rd Canad. Conf. Computational Geometry}, pp. 37--40,
  1991.

\bibitem{RamRamTou-CCCG-95}
S.~Ramaswami, P.~Ramos, and G.~Toussaint.
\newblock {Converting triangulations to quadrangulations}.
\newblock {\em Proc. 7th Canad. Conf. Computational Geometry}, pp. 297--302.
  Centre de recherche en g{\'e}omatique, Universit{\'e} Laval, August 1995.

\bibitem{RodSul-JDG-87}
B.~Rodin and D.~Sullivan.
\newblock {The convergence of circle packings}.
\newblock {\em J. Differential Geometry} 26:349--360, 1987.

\bibitem{ShiGos-SMA-95}
K.~Shimada and D.~C. Gossard.
\newblock {Bubble mesh: automated triangular meshing of non-manifold geometry
  by sphere packing}.
\newblock {\em Proc. 3rd Symp. Solid Modeling {\&} Applications}, pp. 409--419.
  ACM, May 1995.

\bibitem{Ste-CMFT-97}
K.~Stephenson.
\newblock {Approximation of conformal structures via circle packing}.
\newblock {\em Proc. Computational Methods in Function Theory}, 1997,
  \url{http://www.math.utk.edu/~kens/ACS/ACS-revised.ps.gz}.

\end{thebibliography}
\end{document}